\documentstyle[aps, prl, twocolumn]{revtex} 
\begin{document}
\draft

\title{Collective versus local measurements 
on two parallel or antiparallel spins}

\author{S. Massar}
\address{Service de Physique Th\'eorique, Universit\'e Libre de Bruxelles, 
  CP 225, Bvd du Triomphe, B1050 Bruxelles, Belgium}
\date{\today}

\maketitle

\begin{abstract}
We give a complete analysis of covariant measurements on two spins. We 
consider the cases of two parallel  and two antiparallel spins, and
we consider both collective measurements on the two spins, and measurements 
which require only Local Quantum Operations and Classical 
Communication (LOCC). In all cases we obtain the optimal measurements
for arbitrary fidelities. In particular 
we show that if the aim is determine as well as
possible the direction in which the spins are pointing, it is best to
carry out measurements on antiparallel spins (as already shown by
Gisin and Popescu), second best to carry out 
measurements on parallel spins and worst to be restricted to LOCC
measurements. If the the aim is to determine as well as possible a
direction orthogonal to that in which the spins are pointing, it is
best to carry out measurements on parallel spins, whereas measurements 
on antiparallel spins and LOCC measurements are both less good but
equivalent. 
\end{abstract}


\vspace{+0.7cm}

One of the central problems of quantum measurements is how to best
estimate the state of an unknown quantum system. This problem has been 
addressed by many authors, using many different approaches, see
\cite{Helstrom}\cite{Holevo} for reviews.

In the present paper we take a new look at a particular example in
which the task is to determine the direction of polarization of
two identical spin 1/2 particles. We suppose that the polarization direction
is completely unknown, ie. is uniformly distributed on the sphere. 
This problem, generalized to the case of an arbitrary number N of spin 
1/2 has already been studied by several
authors\cite{Holevo,MassarPopescu,DBE,LPT}.
The particular case of 2 spins has the advantage of being
sufficiently simple that a complete
solution can be obtained. Furthermore it allows for several
twists where particular features of quantum mechanics related to
entanglement reveal themselves.

The first twist on the original problem 
was suggested by Peres and Wootters
\cite{PeresWootters}
who asked whether there is a difference between collective, as
compared to local, 
measurements on two particles. Technically, 
in the first case one allows arbitrary quantum operations on both
spins,
whereas in
the second case one restricts oneself to Local Quantum
Operations on each particle and Classical Communication between the
particles (LOCC).

Peres and Wootters gave numerical
evidence that even if two particles are in the same 
state, collective measurements can be better than measurements using
LOCC. An analytical proof was given in \cite{MassarPopescu} in the
case of two identical spin 1/2 whose polarization direction is
uniformly distributed on the sphere, at least in the case where there
are only a finite number of rounds of communication between the two
parties. A remarkable example was
exhibited in \cite{BDFMRSS} which consists of 
a basis of separable states, ie. states
that can be prepared using LOCC, but which nevertheless cannot be
distinguished unambiguously 
using LOCC. In \cite{BDFMRSS} the phrase ``non locality
without entanglement'' was coined for this property. Another result 
is that of \cite{GillMassar} 
where it was shown that in the limit 
of an infinite number of identical spin 1/2, LOCC measurements perform as well
as collective measurements if the spins are in a pure state, but not
as well if the spins are in a mixed state.

A second twist on the original scenario was recently proposed by
Gisin and Popescu \cite{GisinPopescu} who considered the case of two
antiparallel spins. They showed that one can determine better the
direction of polarization of two antiparallel spins than of two
parallel spins. 
Gisin and Popescu's  result is related to ``non locality without
entanglement''  because if
LOCC measurements are carried out on the two spins, it does not make
any difference whether the spins are parallel or antiparallel. Thus
collective measurements on two antiparallel spins is an example of non
locality without entanglement. Mathematically the passage from two
parallel to two antiparallel spins, that is the flip of one of the
spins, is the same operation that Peres used to distinguish whether
two states are entangled or not\cite{Peres}.

The present paper aims at providing an exhaustive analysis of
measurements on two spins 1/2 particles in the three cases of
collective measurements on parallel spins, collective measurements on
antiparallel spins, and LOCC measurements. The spin flip operation
will play a central role in this analysis because it will allow us to
treat all three cases in the same framework.
Our analysis allows one to
find the optimal measurements for arbitrary fidelities. As an
illustration we consider two such fidelities.

The first fidelity 
is  $f=(1 + \cos \theta)/2 $
where $\theta$ is the angle between the
direction in which the spins are polarized and 
the direction in which one guesses that they are polarized. In
this case we recover the results of
\cite{Holevo}\cite{MassarPopescu} 
that if the spins are parallel 
the maximal average fidelity is $f=0.75$. 
If the spins are
antiparallel we show that the maximal fidelity is $f=0.788$. This
fidelity was already obtained in \cite{GisinPopescu} but it was not
known whether it is optimal. Finally we shall show that 
if one restricts oneself to
LOCC measurements, then the maximal fidelity is
$f=0.736$, which is 1.4 \% lower than for measurements on parallel spins.
That this is the optimal value for LOCC measurements was already found by 
 D.G. Fischer, S.H. Kienle, and M. Freyberger \cite{X}.
Thus even in the limit of an infinite number of
rounds of communication collective measurements are better than LOCC
measurements in the case of two parallel spins. 

The second fidelity is
$f=1 - cos^2\theta$. In this case it is most advantageous to guess a
direction orthogonal to that in which the spins are pointing ($\theta
= \pi/2$) and most disadvantageous to guess in the direction in which the
spins are pointing ($\theta =0$) or in the orthogonal direction
($\theta = \pi$). Geometrically this can be rephrased as a situation
in which the spins encode the orientation of a plane by pointing in
the direction normal to the plane and the aim is to find a vector
lying in the plane. In this case the highest  fidelity $f=0.8$ is
 obtained when the spins are parallel.  Antiparallel spins or 
LOCC
measurements both give the same optimal fidelity $f=0.733$.

We now turn to the proof of these results.
Essential to our analysis will be the spin flip operation which we
denote by ${\tilde{\ }}$. For a single spin 1/2 it
takes the form
\begin{equation}
\rho = {I \over 2} + \vec \alpha \cdot \vec \sigma
\quad \to \quad 
\rho{\tilde{\ }} = {I \over 2} - \vec \alpha \cdot \vec \sigma 
\end{equation}
where $I$ is the identity operator and $\sigma_i$ the Pauli spin operators.
In the case of two spins, we will be interested in the operation,
denoted ${\tilde{\ }^2}$ which 
flips only the second spin. If we write the state as
\begin{equation}
\rho = {I \over 4} + \vec \alpha \cdot \vec \sigma \otimes {I\over 2} +
\vec \beta \cdot {I \over 2} \otimes \vec \sigma  + \sum_{i,j}
\gamma_{ij} \sigma_i \otimes \sigma_j \ ,
\end{equation}
then $\rho{\tilde{\ }^2}$ is given by
 \begin{equation}
\rho{\tilde{\ }^2} = 
{I \over 4} + \vec \alpha \cdot \vec \sigma \otimes {I \over 2} -
\vec \beta \cdot {I \over 2} \otimes \vec \sigma  - \sum_{i,j}
\gamma_{ij} \sigma_i \otimes \sigma_j
\end{equation}
The ${\tilde{\ }^2}$ operation is equivalent, up to a unitary
operation acting on particle 2 only, to the partial transpose
introduced in \cite{Peres}.

As an application of the ${\tilde{\ }^2}$ operation consider the
state of two parallel spin 1/2 particles both pointing in the $\vec m$
direction
\begin{eqnarray}
\rho(\vec m,\vec m)&=& |\uparrow_{\vec m}\rangle \langle \uparrow_{\vec m}|
\otimes |\uparrow_{\vec m}\rangle \langle\uparrow_{\vec m}| \nonumber\\ 
&=& {I \over 4} + \sum_{i} m_i (\sigma_i \otimes {I\over 2} +
{I\over 2} \otimes \sigma_i) 
 + \sum_{i,j}
m_i m_j \sigma_i \otimes \sigma_j \nonumber\\
\end{eqnarray}
and the state of two antiparallel spins
\begin{equation}
\rho({\vec m},-{\vec m})=|\uparrow_{\vec m}\rangle \langle \uparrow_{\vec m}|
\otimes |\uparrow_{-\vec m}\rangle \langle\uparrow_{-\vec m}|
 \ .
\end{equation}
We have the relation
\begin{equation}
\rho({\vec m},{\vec m})= \rho({\vec m},-{\vec m}) {\tilde{\ }^2} \ .
\end{equation}

We can also consider the dual of the ${\tilde{\ }^2}$ operation, that
is how it acts on operators. Suppose that $\rho$ is a state and $a$ an 
operator, then $a{\tilde{\ }^2}$ is defined by the relation
\begin{equation}
Tr \ a \rho{\tilde{\ }^2} = Tr\  a{\tilde{\ }^2} \rho\ .
\label{dual}
\end{equation}
One finds that it takes exactly the same form for
operators as it does for states. If the operator $a$ is expressed as
\begin{equation}
a= w {I } + \vec x \cdot \vec \sigma \otimes {I }
+ \vec y \cdot {I } \otimes \vec \sigma  + \sum_{i,j}
z_{ij} \sigma_i \otimes \sigma_j \ ,
\end{equation}
then the operator $a{\tilde{\ }^2}$ takes the form
\begin{equation}
a{\tilde{\ }^2} = w {I } + \vec x \cdot \vec \sigma \otimes {I }
- \vec y \cdot {I } \otimes \vec \sigma  - \sum_{i,j}
z_{ij} \sigma_i \otimes \sigma_j \ .
\end{equation}

The ${\tilde{\ }^2}$ operation for operators 
allows us to put a restriction on the 
Positive Operator Valued Measures
(POVM) acting on the space of two spin 1/2 particles
that can be realized by local Quantum Operations and Classical
Communication (LOCC). Indeed it was shown in \cite{BDFMRSS} that such a POVM,
defined by its elements
$ a_i \geq 0$, $ \sum_i a_i = 1$, must obey
$a_i{\tilde{\ }^2} \geq 0$ for all $i$.

To proceed with the proof,
consider a set of operators $a_i$ that sum to the identity $\sum_i 
a_i = 1$. We are interested in the following 3 positivity
conditions on $a_i$:
\begin{enumerate}
\item $a_i \geq 0$. In this case the $a_i$ constitute a POVM. 
The probability of getting outcome $i$ if the state is $
\rho({\vec m},{\vec m})$ is $P_\parallel(i|{\vec m}) = Tr 
\rho({\vec m},{\vec m}) a_i$.
\item $a_i{\tilde{\ }^2} \geq 0$. In this case the $a_i{\tilde{\ }^2}$ 
  constitute a POVM. The probability of getting outcome $i$ if the state is $
\rho({\vec m},-{\vec m})$ is $P_\perp(i|{\vec m}) = Tr 
\rho({\vec m},-{\vec m}) a_i{\tilde{\ }^2}$. Using equation (\ref{dual}) we have
$ P_\perp(i|{\vec m})= Tr \rho({\vec m},{\vec m}) a_i$.
\item  $a_i \geq 0$ and $a_i{\tilde{\ }^2} \geq 0$. In this case both
  $a_i$ and $a_i{\tilde{\ }^2}$ constitute a measurement which can be
  realized by LOCC. The
  probability $P_\parallel(i|{\vec m})= 
Tr \rho({\vec m},{\vec m}) a_i$
of obtaining outcome $i$ if the spins are  parallel
 and the measurement is $a_i$ equals the
  probability $ P_\perp(i|{\vec m})= 
Tr \rho({\vec m},{\vec -m}) a_i{\tilde{\ }^2}$
of of obtaining outcome $i$ if the spins are
  antiparallel 
and the measurement is $a_i{\tilde{\ }^2}$.
The equality of
$P_\parallel(i|{\vec m})$ and $ P_\perp(i|{\vec m})$ shows that in this
case there is no difference between making measurements on parallel
and antiparallel spins.
\end{enumerate}

Thus the ${\tilde{\ }^2}$ operation relates 
measurements on parallel spins (given by $P_\parallel(i|{\vec m})$),
measurements on antiparallel spins (given by $P_\perp(i|{\vec m})$), and 
measurements that can be realized by LOCC. The central idea is that by 
using
the ${\tilde{\ }^2}$ operation all these quantities can be
expressed in terms of the same trace $Tr\ \rho({\vec m},{\vec m})
a_i$, but with operators $a_i$ which obey the different positivity conditions
enumerated above.

To further explicitise these different positivity conditions we shall 
suppose that the aim of the
measurement is to distinguish along which direction the spins are
pointing. We can then label the POVM elements $a_{\vec n}$ 
by the direction $\vec
n$ along which one guesses the spins are pointing.  
Furthermore we shall suppose that the
spins are polarized in a random direction uniformly
distributed on the sphere. We can then, without loss of generality
\cite{Holevo},
suppose that we are dealing with covariant measurements, that is
measurements for which the guessed direction $\vec n$ spans the whole
sphere and which satisfy
\begin{equation}
Tr a_{\vec n} \rho({\vec m},{\vec m}) = 
Tr a_{R(\vec n)} \rho(R({\vec m}),R({\vec m})) 
\end{equation}
where $R$ is an arbitrary rotation, ie. an element of $SO(3)$.
Using the fact that $\rho(R({\vec m}),R({\vec m})) =
R\otimes R \rho({\vec m},{\vec m}) R^\dagger \otimes R^\dagger$ where
$R$ 
is the corresponding element of $SU(2)$, we have
\begin{equation}
a_{R(\vec n)} = R\otimes R a_{\vec n} R^\dagger \otimes R^\dagger \ .
\label{cov}
\end{equation}
We can also without loss of generality suppose that the measurement is 
symmetric with respect to 
 interchanges the two spins. This implies that
\begin{equation}
a_{\vec n} = w {I } + \vec x \cdot (\vec \sigma \otimes {I }
+ {I } \otimes \vec \sigma )  + \sum_{i,j}
z_{ij} \sigma_i \otimes \sigma_j \label{symm}
\end{equation}
with $z_{ij}$ a symmetric matrix.

The covariance condition (\ref{cov}) 
implies a considerable simplification on the coefficients $w, \vec x, 
z_{ij}$ in (\ref{symm}).
Consider the POVM element $a_{\vec z}$
corresponding to guessing the spins are polarized along the $+z$
direction.
Let $R_{\phi,z}$ be a rotation of angle $\phi$ around the $z$ axis. We
have
$a_{\vec z} = R_{\phi,z}\otimes  R_{\phi,z} a_{\vec z} 
R_{\phi,z}^\dagger\otimes R_{\phi,z}^\dagger$ for all
$\phi$. Using (\ref{symm}), this implies that $a_{\vec z}$ has the form
\begin{eqnarray}
a_{\vec z} &=& w {I } + {\alpha } (\sigma_z \otimes {I } +
{I } \otimes \sigma_z ) + \beta \sigma_z \otimes \sigma_z
\nonumber\\
& &
+ \gamma ( \sigma_x \otimes \sigma_x + \sigma_y \otimes \sigma_y)
\label{cov1}
\end{eqnarray}
where $\alpha$, $\beta$, $\gamma$ are three real numbers. 

A final simplification results if we recall that the operators
$a_{\vec n}$
must sum to the identity:
\begin{equation} \int d\vec n\  a_{\vec n} = 
\int_{SU(2)} d R \ R\otimes R a_{+\vec z} R^\dagger \otimes R^\dagger = I\ .
\end{equation}
Using (\ref{cov1}) this implies that
\begin{eqnarray}
a_{\vec z} &=& {I } + {\alpha} (\sigma_z \otimes {I } +
{I } \otimes \sigma_z ) 
\nonumber\\
& &  + \gamma (  2\sigma_z \otimes \sigma_z
- \sigma_x \otimes \sigma_x - \sigma_y \otimes \sigma_y)
\label{cov2}
\end{eqnarray}
which only depends on two parameters $\alpha$ and $\gamma$.

It is now easy to compute the restriction on the two parameters
$\alpha$ and $\gamma$ which result from each of the three positivity
conditions enumerated above:
\begin{eqnarray}
a_{\vec n} \geq 0 
\ &\Rightarrow& \ 
\gamma \leq 1 
\ ,\ 
1+ \alpha + \gamma /2 \geq 0 
\ , \nonumber\\
& &\ 
1 - \alpha +  \gamma /2 \geq 0  
\label{cond1}
\\
a_{\vec n}{\tilde{\ }^2} \geq 0 \ ,
\ &\Rightarrow& \
\gamma \leq 2 
\ ,\ 
1 +  \gamma - \alpha^2 \geq 0 \ ,
\label{cond2}
\\
a_{\vec n}\ \mbox{and}\ a_{\vec n}{\tilde{\ }^2} \geq 0
\ &\Rightarrow& \ 
\gamma \leq 1 
\ ,\ 
1 +  \gamma - \alpha^2 \geq 0 \ .
\label{cond3}
\end{eqnarray}

These constitute convex sets. The extremal
points of these convex sets will be the optimal
measurements. To understand what extremal point corresponds to what
optimal measurement we introduce a fidelity function $f$. We now study 
different fidelity functions.

The covariance
of the measurement set up implies that $f$ is a function only of the
angle between the direction in which the spins are polarized $\vec m$
and the direction guessed by the POVM element $a_{\vec n}$. Thus in
the case of outcome $+z$, $f$ is a function of $m_z$ only.
It is convenient to expand $f$ in Legendre polynomials
\begin{eqnarray}
f(+z,\vec m)&=&
\sum_{n=0}^\infty f_n P_n(m_z) 
\nonumber\\
&=& f_0 + f_1 m_z + f_2 {3 m_z^2 - 
  1 \over 2} + \cdots \ .
\end{eqnarray}

To compute the average fidelity we need the probability of each outcome.
Suppose that the spins point in direction $\vec m$ and that the
measurement outcome is $+z$. This occurs with probability
\begin{eqnarray}
P(+z | \vec m)=Tr \rho(\vec m ,\vec m) a_{\vec z}  
&=& 1 +  \alpha m_z + {\gamma \over 2 } {3 m_z^2 - 1 \over 2} \ .
\end{eqnarray}
The average fidelity is therefore
\begin{eqnarray}
F &=& \int_{-1}^{+1} {d m_z \over 2} \int_0^{2\pi} {d\phi \over 2  \pi}
f(+z,\vec m) P(+z|\vec m)\nonumber\\
&=& f_0 +   {\alpha \over 3}f_1  +  {\gamma \over 10 }f_2\ .
\label{fid}
\end{eqnarray}
Thus only the first three coefficients enter into the average
fidelity. (In the case of covariant measurements on
$N$ parallel spins only the $N+1$ first
coefficients of the expansion of $f$ will enter into the average fidelity).

Using eqs. (\ref{cond1}, \ref{cond2}, \ref{cond3}) and (\ref{fid}) it is 
straightforward to
find the optimal measurement for an arbitrary fidelity function in the 
case of parallel spins, antiparallel spins and LOCC measurements.
As a first illustration, 
let us consider the example studied in \cite{Holevo} and
\cite{MassarPopescu} in which the fidelity has the form $f(\vec n| \vec m) = 
|\langle \uparrow_{\vec m}|\uparrow_{\vec n}\rangle|^2 = (1 + \cos
  \theta )/ 2$. Thus in this example $f_0=1/2$, $f_1=1/2$, $f_2=0$
and therefore $F = 1/2 + \alpha /6$. In this
case the largest fidelity is obtained by taking for $\alpha$ the
largest possible value. In the case of two parallel spins the largest
possible value of $\alpha$ is $\alpha_{max}=3/2$ corresponding to
$F_{\parallel}= 3/4 = 0.75$, a result already obtained in \cite{Holevo} and
\cite{MassarPopescu}. In the case of two antiparallel spins  
$\alpha_{max}=\sqrt{3}$ corresponding to
$F_{\perp}= 1/2 + 1/(2\sqrt{3}) \simeq 0.788$, 
a result already obtained in \cite{GisinPopescu}.
In the case of measurements carried out using only LOCC,   
$\alpha_{max}=\sqrt{2}$ corresponding to
$F_{LOCC}= 1/2 + 1/(3\sqrt{2}) \simeq 0.736$. Thus for this fidelity
collective measurements on antiparallel spins are better than
collective measurements on parallel spins which are themselves better
than LOCC measurements on parallel (or antiparallel) spins.

Note that if the spins are parallel or antiparallel, optimal
measurements that use a 1 dimensional ancilla (which could be the
singlet state) have been exhibited in \cite{MassarPopescu} 
and \cite{GisinPopescu}. In the case of LOCC measurements it is easy
to check that a simple optimal
strategy consists in Alice making a von Neumann measurement of spin
along some direction $\vec a$ and Bob making a von Neumann measurement 
of spin along an orthogonal direction $\vec b$ ($\vec a \cdot \vec b
=0$). 
Denote by $\vec \alpha 
= \pm \vec a$ and $\vec \beta 
= \pm \vec b$ the results of the two measurements. Then the guessed
direction is the bisectrix $\vec \alpha + \vec \beta$ of the two
results.
Thus in all cases 
the optimal measurements can be implemented using rather simple
strategies which necessitate low dimensional ancillas. This should be
compared with the covariant measurements which, although useful for
the theoretical analysis, require infinite dimensional ancillas. 

As a second illustration consider the case where the fidelity is 
$f=\sin ^2 \theta = 1 - \cos^2 \theta$. 
In this case $f_0=2/3$, $f_1=0$, $f_2=-2/3$, hence 
$F = 2/3 - \gamma / 15$ and the best measurement is that which has the
smallest value of $\gamma$. In the case of collective measurements on
parallel spins the smallest value is $\gamma_{min}=-2$ yielding a
fidelity $F_\parallel = 4/5=0.8$. For collective measurements on 
antiparallel spins or LOCC measurements the minimum value is
$\gamma_{min}=-1$  
yielding an optimal fidelity $F_{\perp , LOCC}= 11/15\simeq
0.733$. 
In this case measurements on two parallel spins are better than
measurements on antiparallel spins or LOCC measurements which are both 
equivalent.
Thus for some fidelities measurements on parallel spins are 
better, for other fidelities measurements on antiparallel spins are
better, and in all cases LOCC measurements are the worst.

In conclusion the present article gives explicitly the optimal
measurements and optimal fidelity for all possible fidelity functions
in the cases of parallel spins, antiparallel spins, and LOCC
measurements. This provides an interesting target for future
experiments since it provides a criterion for putting
 ``non locality without entanglement'' into evidence.

\vspace{0.5cm}

I would like to thank N. Gisin, N. Linden and S. Popescu for
stimulating and helpful discussions, and Dietmar Fisher for pointing
out his recent work to me.
I am a research associate of the 
Belgian National Fund for Scientific Research.

\end{document}